**This is the accepted manuscript (postprint) of the following article:**
Majid Hosseinzadeh, Erfan Salahinejad, *Comparative analysis of electrodeposited Pt, Ru and Pt-Ru overlays for high-temperature oxidation protection*, Surface and Coatings Technology, 496 (2025) 131685.
https://doi.org/10.1016/j.surfcoat.2024.131685

# Comparative Analysis of Electrodeposited Pt, Ru and Pt-Ru Overlays for High-Temperature Oxidation Protection

Majid Hosseinzadeh, Erfan Salahinejad*

*Faculty of Materials Science and Engineering, K.N. Toosi University of Technology, Tehran, Iran*

## Abstract

Platinum (Pt) and ruthenium (Ru), both members of the platinum-group metals (PGMs), are renowned for their exceptional resistance to corrosion, oxidation, and high temperatures, making them promising candidates for advanced high-temperature applications. This study investigates the direct current (DC) electrodeposition of Pt, Ru, and a binary Pt-Ru alloy onto NiCoCrAlYTa-coated single-crystal superalloy CMSX-4, along with their vacuum annealing and respective effects on the isothermal oxidation behavior of the system at 1100 °C. All the electrodeposited overlays demonstrated substantial enhancement in oxidation resistance. However, Pt exhibited the highest protection efficiency, Ru the least, and the Pt-Ru alloy provided an intermediate level of performance. Microscopic and X-ray diffraction analyses revealed that the competitive formation of protective α-$Al_2O_3$ and spinel $NiAl_2O_4$ phases on the coated surfaces played a crucial role in determining the oxidation resistance, driven by atomic interactions between the elements in the NiCoCrAlYTa bond coat and the overlay metals. Despite Ru's relatively lower oxidation resistance compared to Pt, its significantly lower cost offers potential advantages in cost-sensitive, high-temperature applications. These findings provide valuable insights into optimizing Pt-group metal coatings for durability in high-performance systems.

* Corresponding Author's Email Address: <salahinejad@kntu.ac.ir>

## 1. Introduction

The aspiration of industries that use gas turbines, including power generation, oil and gas, marine, and aerospace, has always been to develop and use thermal barrier coatings (TBCs) that provide thermal insulation. In addition to TBCs that are primarily composed of ceramic materials, aluminide and MCrAlY (M stands for Ni, Co, or a combination of both) are extensively employed as bond coatings between the substrate and TBCs or as overlay coatings to protect the underlying metal from extreme heat. The deterioration of these metallic coatings, particularly due to oxidation, is a significant cause of the component failure [1-3]. The high-temperature resistance of these coatings is further enhanced through different strategies, including incorporation with platinum (Pt). Typically, Pt-modified aluminide diffusion coatings are used in certain industrial applications to offer protection efficiencies exceeding those of Pt-free aluminide coatings [4-6]. However, the exorbitant cost of Pt imposes significant constraints on its use, particularly for MCrAlY coatings that are already expensive, limiting its widespread applicability.

Ruthenium (Ru), another member of the Pt-group metals, exhibits thermodynamic properties similar to Pt, but costs only about a quarter of Pt. The increase in the resistance of aluminide and MCrAlY coatings to oxidation via Pt [7-9] and Ru [10-12] modifications has been distinctly reported. However, there are no comparative studies on the protection efficiency of these two additives, to our knowledge. In addition, the chemical and electrochemical similarity of these two elements allows for the formation of their solid solution alloys across a wide range of composition, as indicated by their binary phase diagram [13]. While Pt-Ru alloys have been reported in catalytic applications such as direct methanol fuel cells [14-16], they have not been studied in high-temperature applications.

This work is focused on the comparative high-temperature oxidation characterization of MCrAlY coatings modified with Pt, Ru, and Pt-Ru overlays to explore the potential of Ru as an entire or partial replacement for expensive Pt in gas turbines. It is noteworthy that the deposition and subsequent effects of Ru overlays on both aluminide and MCrAlY coatings have not been reported before. In this study, as in most previous studies, electrodeposition is used to apply these layers due to the high melting points of the elements as well as high density results, equipment simplicity, and affordability associated with electrodeposition [17-19].

## 2. Experimental procedure

### *2.1. Sample preparation and characterization*

A Ni-based single-crystal superalloy-CMSX-4 (nominal composition in wt%: 60.9Ni-9.5Co-6.5Ta-6.4Cr-6.3W-5.7Al-2.9Re-1.0Ti-0.6Mo-0.1Hf) was used as the substrate material, while disks of 10 mm in diameter and 1 mm in thickness were cut from rods. A NiCoCrAlYTa layer of almost 120±10 μm in thickness was deposited on the substrate by a high velocity oxygen fuel (HVOF) technique from Amdry997 powder (Ni-23.0Co-20.0Cr-8.5Al-4.0Ta-0.6Y in wt%). HVOF deposition operation was conducted at a spraying distance of 255 mm, power feed rate of 50 g/min, $O_2$ flow of 800 L/min and $N_2$ flow of 15 L/min. Afterward, Pt, Ru, and Pt-Ru layers were electroplated on the NiCoCrAlYTa layer using electrolyte and electrodeposition conditions listed in Table 1, adapted from Refs. [19-21], [19, 22, 23], and [24-26], respectively. The electroplating cell used had a fixed volume of 250 mL with a mixed metal oxide (MMO) anode whose surface area was three times larger than that of the cathode. The distance between the cathode and anode was kept constant at 4 cm throughout the electroplating process. To maintain a stable pH of the electrolyte, HCl and NaOH were used. The coated samples were vacuum-annealed at 1080 °C for 6 h.

Table 1. Bath composition and electroplating parameters for depositing Pt, Ru, and Pt-Ru coatings.

| **Compositions/Parameters** | **Pt coating** | **Ru coating** | **Pt-Ru coating** |
|---|---|---|---|
| $Pt(NH_3)_2(NO_2)_2$ (mol/L) | 0.03 | - | 0.03 |
| $Na_2CO_3$ (mol/L) | 0.15 | - | 0.15 |
| $NaCH_3COO$ (mol/L) | 0.12 | - | 0.12 |
| $RuCl_3$ (mol/L) | - | 0.02 | 0.02 |
| HCl (mol/L) | - | 0.30 | 0.30 |
| NaCl (mol/L) | - | 0.03 | 0.03 |
| Current density (A/dm$^2$) | 0.5 | 2 | 3 |
| Temperature (ºC) | 90 | 70 | 75 |
| Solution pH | 11 | 1 | 3 |
| Time (min) | 180 | 60 | 90 |

The top and cross section of the coatings were characterized using field-emission electron microscopy (FESEM, MIRA3, TESCAN) equipped with energy dispersive X-ray spectrometry (EDS). Reference lines for EDS line scans were selected from regions of the coatings that visually appeared representative of the overall morphology and had thicknesses close to the calculated average coating thickness. The thickness is expressed as the mean ± standard deviation across three repeated samples, with measurements obtained from three regions per sample. The coatings were also tested for bonding strength according to the ASTM C633 standard [27]. The method involved fastening the coated samples to an uncoated sample using SW 2214 epoxy adhesive (3M Scotch-weld, bond strength~70 MPa), which was applied at 150 °C for 2 h to ensure a strong bond. The samples were then subjected to a tensile load at a rate of 5 mm/min until fracture, and the corresponding fracture stress was recorded with five repetitions.

### *2.2. Analysis of oxidation protection*

Isothermal oxidation tests were carried out in a muffle furnace under static air conditions at 1100 °C up to 100 h. Mass changes of the oxidized samples together with an alumina crucible were measured at regular intervals using an electronic balance with the sensitivity of $10^{-4}$ g at three repetitions. In addition, FESEM-EDS and X-ray diffraction (XRD) analyses were conducted to explore the surface morphology, composition, and phase of the oxidated coatings.

## 3. Results and discussion

### *3.1. Structural and bonding characterization of the annealed samples*

The secondary electron (SE) surface morphology and EDS spectra of the electrodeposited, annealed Pt, Ru, and Pt-Ru coatings are illustrated in Fig. 1. As can be observed, the Pt coating exhibits a globular structure in agreement with Refs. [28-30], while the Ru coating has a cauliflower structure that is compatible with Refs. [22] and the Pt-Ru coating is characterized by a dendritic growth mode that aligns with Ref. [31]. In general, the globular structure results from uniform nucleation across the substrate, leading to consistent growth in all directions. This occurs under charge-transfer controlled deposition, where a high concentration of ions and low overpotential promote stable, compact grain formation as atoms minimize surface energy. This morphology is characteristic of elements that exhibit high deposition efficiency, high reduction potential, and are supplied in sufficient concentrations [32-34]. In contrast, the cauliflower structure arises from diffusion-controlled growth, where the rate of nucleation exceeds the rate of diffusion, resulting in uneven growth and the formation of a more disordered structure [32, 34, 35]. The dendritic morphology of the Pt-Ru coating arises from the simultaneous co-deposition of Pt and Ru, driven by their differing deposition mechanisms. Pt, with charge-transfer-controlled deposition, forms the core structure through compact nucleation due to its higher reduction potential. Ru, governed by diffusion-controlled deposition, contributes to branching by nucleating on the Pt framework. This interplay of mechanisms is characteristic of dendritic structures formed through co-electrodeposition [36-39].

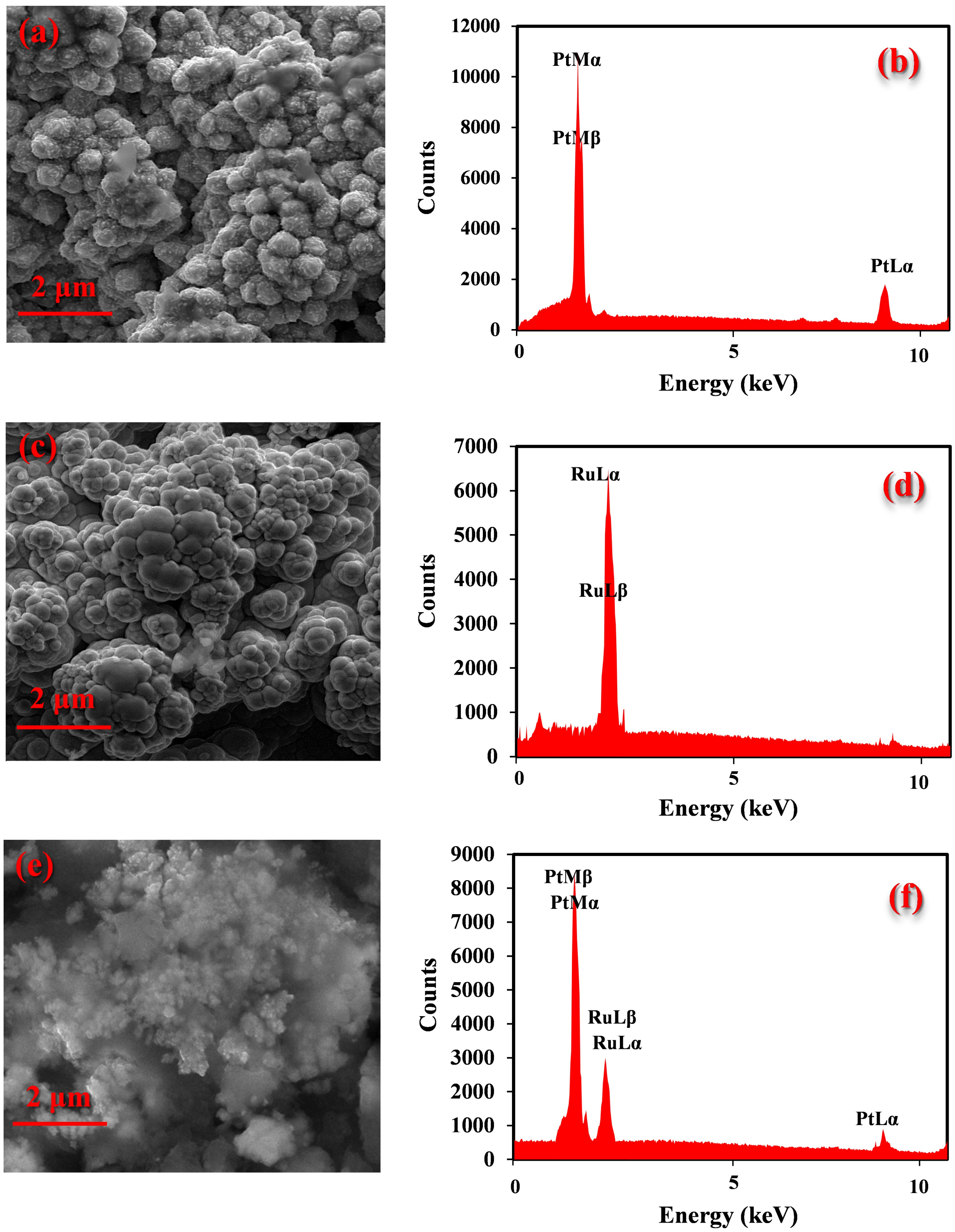

Fig. 1. SE micrographs and EDS spectra of the electroplated pure Pt (a, b), pure Ru (c, d), and Pt-Ru alloy (e, f) coatings.

The chemical composition of the coatings was validated by the EDS analysis, with 66Pt-34Ru (wt%) obtained for the alloy deposit, while the loaded weight ratio of Pt to Ru was 60/40 to provide a single-phase platinum-rich alloy based on the binary phase diagram [40]. However, the actual composition of the alloy deposited by electroplating is determined by not only the nominal ionic composition of the bath, but also the differential electrodeposition rates of elements [25, 41]. Typically, Pt, with its higher reduction potential, tends to deposit more rapidly in diffusion-limited conditions, while Ru, with a lower reduction potential, deposits more slowly under kinetically-limited conditions [39].

The cross-sectional backscattered electron (BSE) images, EDS linear scans, and elemental maps taken from the samples are presented in Figs. 2 and 3. The Pt, Ru, and Pt-Ru coatings exhibit thicknesses of 6.0 ± 0.5, 5.0 ± 1.0, and 5 ± 0.5 µm, respectively. These measurements are consistent with the thicknesses predicted by Faradaic thickness calculations [42], indicating that the deposition processes were well-controlled. The standard deviations in thickness aligns with the morphological structures observed in Fig. 1, influencing the uniformity of the coating thickness. Globular structures, as seen in the Pt coating, typically result in more uniform coatings with lower standard deviations. In contrast, more complex morphologies, such as the cauliflower-like structure of the Ru coating and the dendritic growth observed in the Pt-Ru coating, contribute to greater variability in the coating thickness. This relationship highlights how the morphology influences the uniformity of the coating thickness, with more irregular structures leading to increased thickness variability.

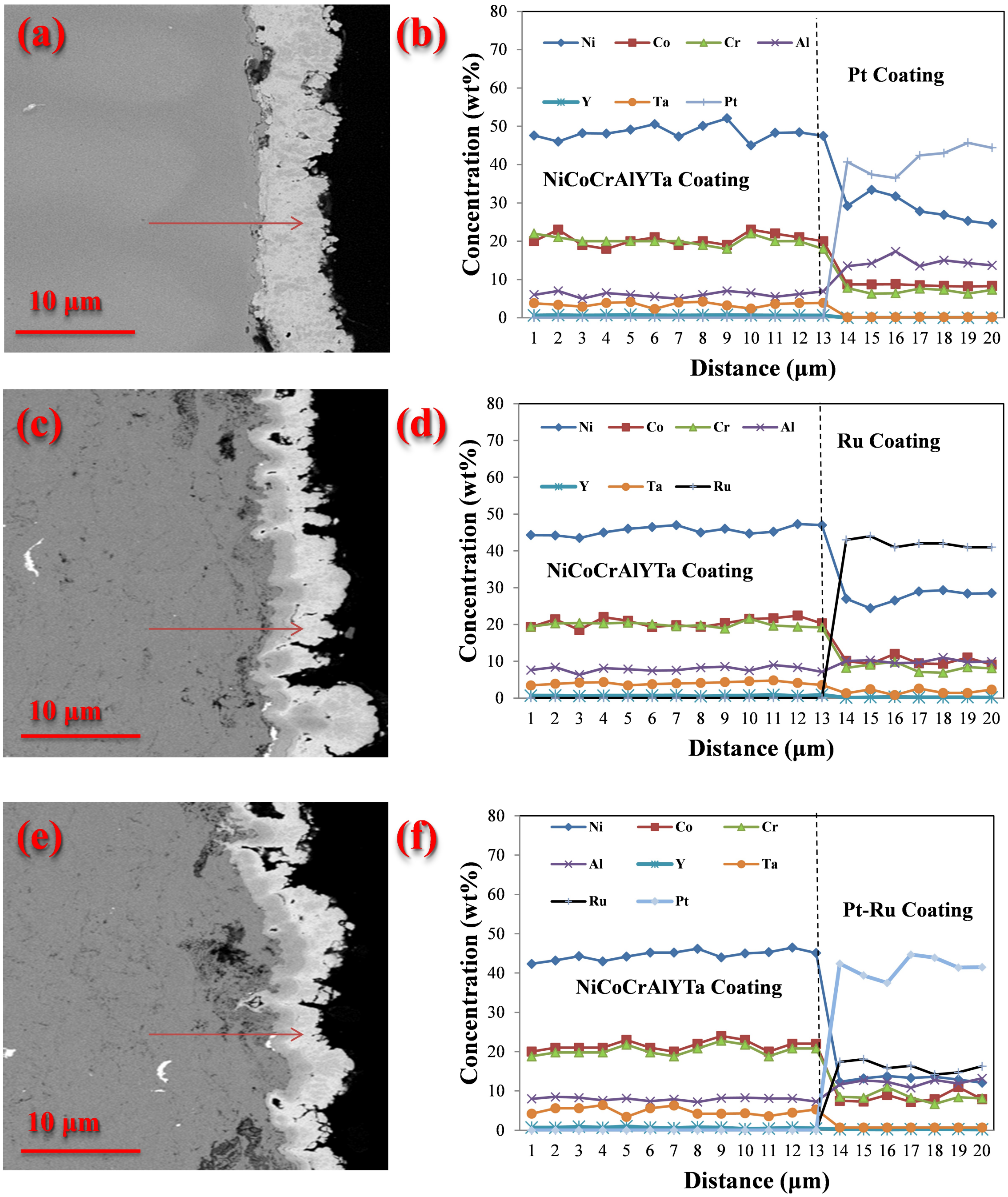

Fig. 2. BSE cross-sectional images and corresponding EDS line scans for the Pt (a, b), Ru (c, d), and Pt-Ru (e, f) coatings.

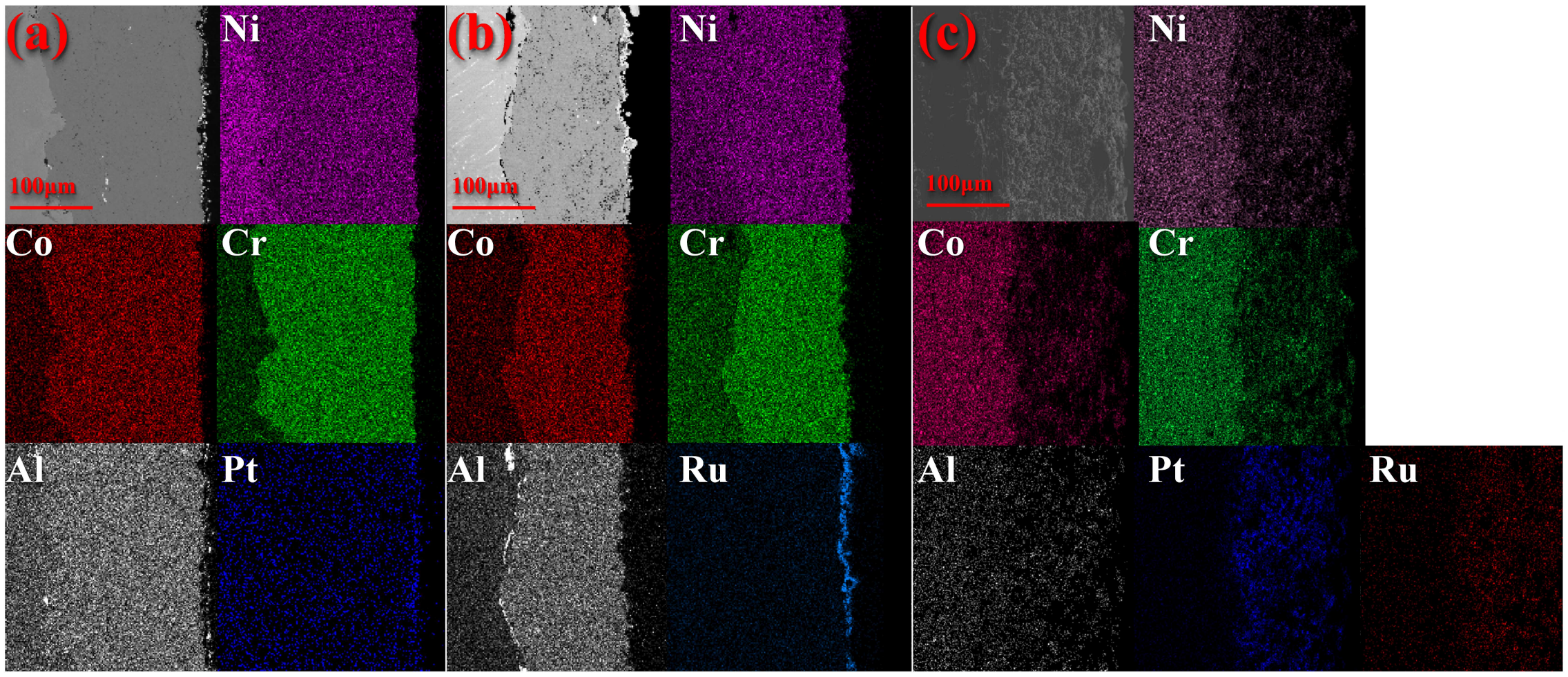

Fig. 3. BSE cross-sectional images and EDS elemental distribution maps for the Pt (a), Ru (b), and Pt-Ru (c) coatings.

The EDS linear scans and elemental maps confirm the relatively uniform distribution of Pt and Ru at the microscale within the coatings. No significant interdiffusion of Pt and Ru into the underlay coat was detected. However, certain constituent elements of the bond coat have outward diffused into the Pt-group overlays, with surface concentrations listed in Table 2. As observed, the surface concentrations of Ni, Co, Cr, and Al on the Pt overlay are higher than those on the Ru coating, while the concentrations of Y and Ta on the Ru coating are higher compared to the Pt overlay. The concentrations on the alloy overlay fall between those observed on the elemental coatings.

Table 2. EDS-extracted concentration of the bond coat elements on the surface of the electrodeposited overlays

| Coatings | Elements | | | | | |
|---|---|---|---|---|---|---|
| | Ni | Co | Cr | Al | Y | Ta |
| **Pt** | 33.4 | 8.6 | 9.1 | 14.1 | 0 | 0.2 |
| **Ru** | 31.5 | 7.8 | 8.3 | 5.5 | 0.2 | 1.7 |
| **Pt-Ru** | 32.1 | 8.2 | 8.6 | 8.4 | 0 | 0.7 |

The distribution of elements in diffusion couples is generally determined by both thermodynamic and kinetic factors in terms of interactions and diffusivities of the elements involved, respectively. The enthalpy of mixing of a pair is closely related to their affinity, which describes how strongly they interact with each other when mixed. A positive enthalpy of mixing generally indicates that phase separation or limited solubility is energetically more favorable than the formation of a stable mixed phase or interdiffusion. A moderately negative enthalpy of mixing usually indicates the formation of a solid solution or interdiffusion, while highly negative enthalpy of mixing typically suggests the formation of new phases like intermetallic compounds with slower diffusivity due to the more rigid and ordered lattice structure [43, 44]. Table 3 lists the enthalpy of mixing for the elements in the bond coat when combined with Pt and Ru. To explain the interdiffusion of Pt and Ru in the bond coat, it is indeed practical to focus primarily on interactions between Pt and Ru with Ni and Co as they are the major components in the bond coat alloy (Ni-23.0Co-20.0Cr-8.5Al-4.0Ta-0.6Y in wt%). The small negative enthalpy of mixing for these pairs indicates a limited affinity between the elements, resulting in the restricted interdiffusion of Pt and Ru into the bond coat. In contrast, Ni and Co exhibit higher interdiffusion into the overlays, as their smaller atomic sizes facilitate greater diffusivity compared to Pt and Ru. The moderately negative enthalpy of mixing for Cr and Al with Pt and Ru explains their interdiffusion into the overlay coatings, with a higher concentration observed in the Pt coating due to its greater affinity for these elements. The moderately negative enthalpy of mixing for Y and Ta with Ru explains their interdiffusion into the Ru overlay, while the highly negative enthalpy of mixing for Y and Ta with Pt is responsible for their blocked interdiffusion into the Pt overlay. This is supported by their binary phase diagrams, which demonstrate the low solubility of Y and Ta in Pt, as well as the formation of multiple intermetallic compounds between these pairs [45].

Table 3. Enthalpy of mixing of the elements existing in the bond coat with Pt and Ru [43, 44]

| **$\Delta H_{mix}$ (kJ/mol)** | | | | | | |
|---|---|---|---|---|---|---|
| | **Ni** | **Co** | **Cr** | **Al** | **Y** | **Ta** |
| **Pt** | -5 | -7 | -24 | -44 | -83 | -66 |
| **Ru** | 0 | -1 | -12 | -21 | -34 | -39 |

The adhesion strength of the Pt, Ru, and Pt-Ru coatings was measured to be 43.2 ± 3.5, 39.1 ± 4.5, and 40.3 ± 4.0 MPa, respectively. While the variations in the bonding strength of the coatings are minimal, they can be explained by considering the observed surface morphology (Fig. 1) and the interdiffusion of the atoms between the electrodeposited coatings and underlying bond coat (Fig. 2). Surface morphology influences bonding strength by affecting the uniformity and compactness of the coating-substrate interface [46, 47]. Typically, a more uniform and compact morphology, such as the Pt's globular structure, enhances bonding strength, while complex, irregular morphologies, like Ru's cauliflower structure, can weaken it. Pt-Ru, exhibiting dendritic growth, combines the characteristics of both materials, resulting in intermediate bonding strength. The bonding strength can also be influenced by thermodynamic and kinetic interactions between the elements in the layers [48-50]. Pt shows stronger affinity and interdiffusion with the principal elements of the underlying coating, which is favorable for adhesion, while the Ru's lower affinity and diffusion could limit its ability to bond effectively to the substrate. The moderate behavior of Pt-Ru likely leads to an intermediate bonding strength due to the varied behaviors of Pt and Ru within the coating.

### *3.2. Isothermal oxidation characterization of the samples*

The 1100 °C isothermal oxidation kinetic curves for the samples are illustrated in Fig. 4. All the samples exhibit a sharp weight gain within the first hour of exposure due to the rapid formation of oxide scales, a common behavior during high-temperature oxidation where the formation of an oxide layer protects the underlying metal from further oxidation [51, 52]. As can be also seen, the employment of the electrodeposited overlayers has significantly improved oxidization resistance when compared with the control sample (bare NiCoCrAlYTa). Comparing the electrodeposition-coated samples, the Pt-coated sample exhibits the least weight gain, the Ru-coated sample the highest, and the Pt/Ru-coated sample an intermediate amount over the entire oxidation period. This suggests that Pt promotes the formation of a more stable and continuous oxide layer compared to Ru, which leads to better protection of the underlying metal substrate.

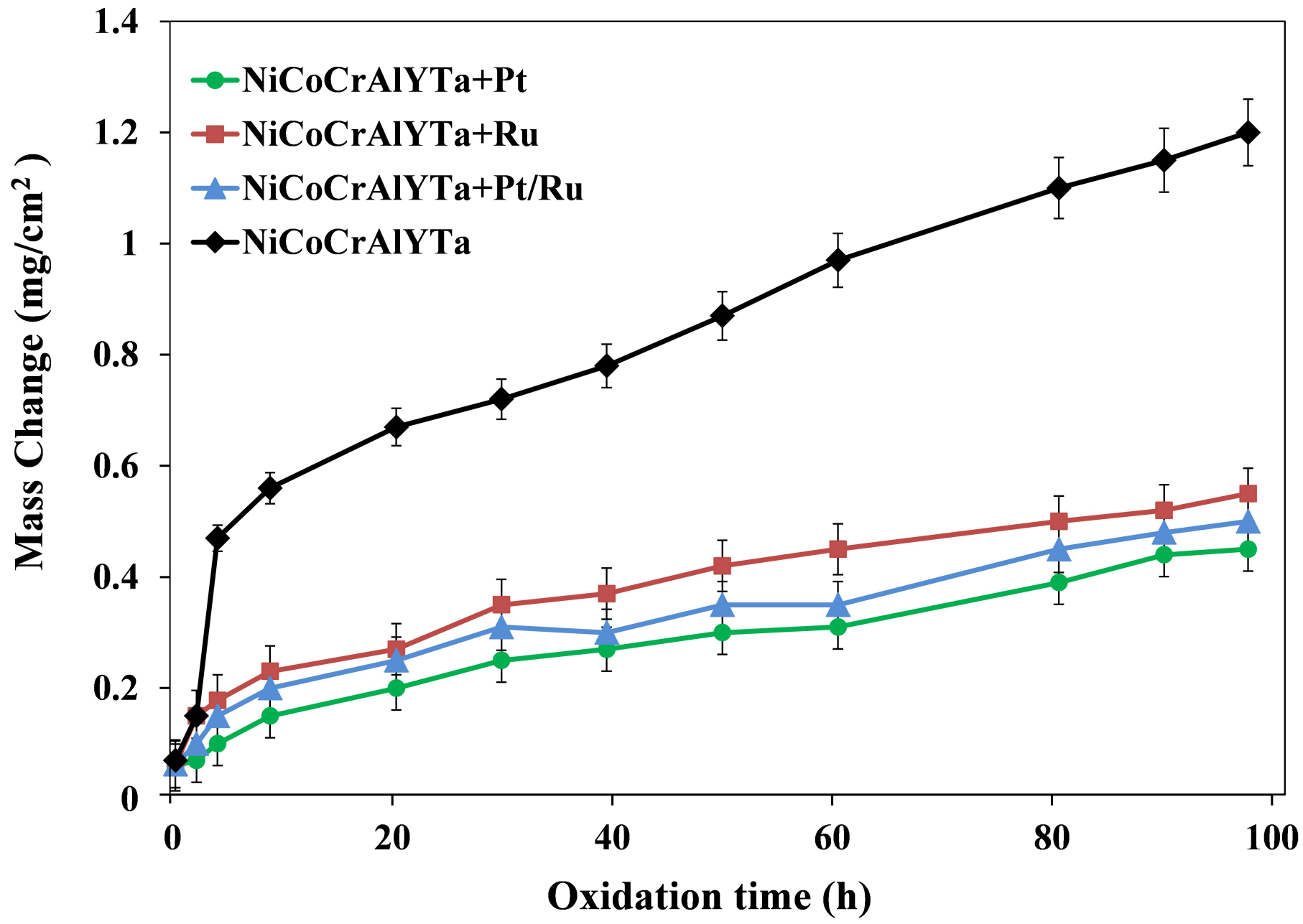

Fig. 4. Isothermal oxidation curves of the NiCoCrAlYTa+Pt, NiCoCrAlYTa+Ru, and NiCoCrAlYTa+Pt/Ru samples.

Fig. 5 depicts the XRD patterns of the samples after isothermal oxidation at 1100 °C for 100 h. Based on the peak analysis, the dominant oxide scale formed on all the coating surfaces is $Al_2O_3$ with weak diffraction signals from the spinel $NiAl_2O_4$ phase. Also, the Pt, Ru, and Pt/Ru-coated samples contain γ-Ni(Pt) / γ'-$(Ni,Pt)_3Al$, γ-Ni(Ru) / γ'-$(Ni,Ru)_3Al$, and γ-Ni(Pt,Ru) / γ'-$(Ni,Pt,Ru)_3Al$, respectively. The relative number and intensity of the $Al_2O_3$ peaks rank as Pt > Pt-Ru > Ru, which aligns well with the elemental distribution in the overlay coatings (Fig. 2) and the oxidation resistance ranking (Fig. 4), as $Al_2O_3$ serves as the essential phase providing oxidation protection in the thermally-grown oxide (TGO) layer [7, 53, 54]. In contrast, diffraction signals for the spinel $NiAl_2O_4$ phase rank as Ru > Pt-Ru > Pt, which is in consistent with the observed oxidation behavior, as this phase deteriorates the oxidation protection of the TGO due to a higher growth rate compared to α-$Al_2O_3$ [52, 55, 56].

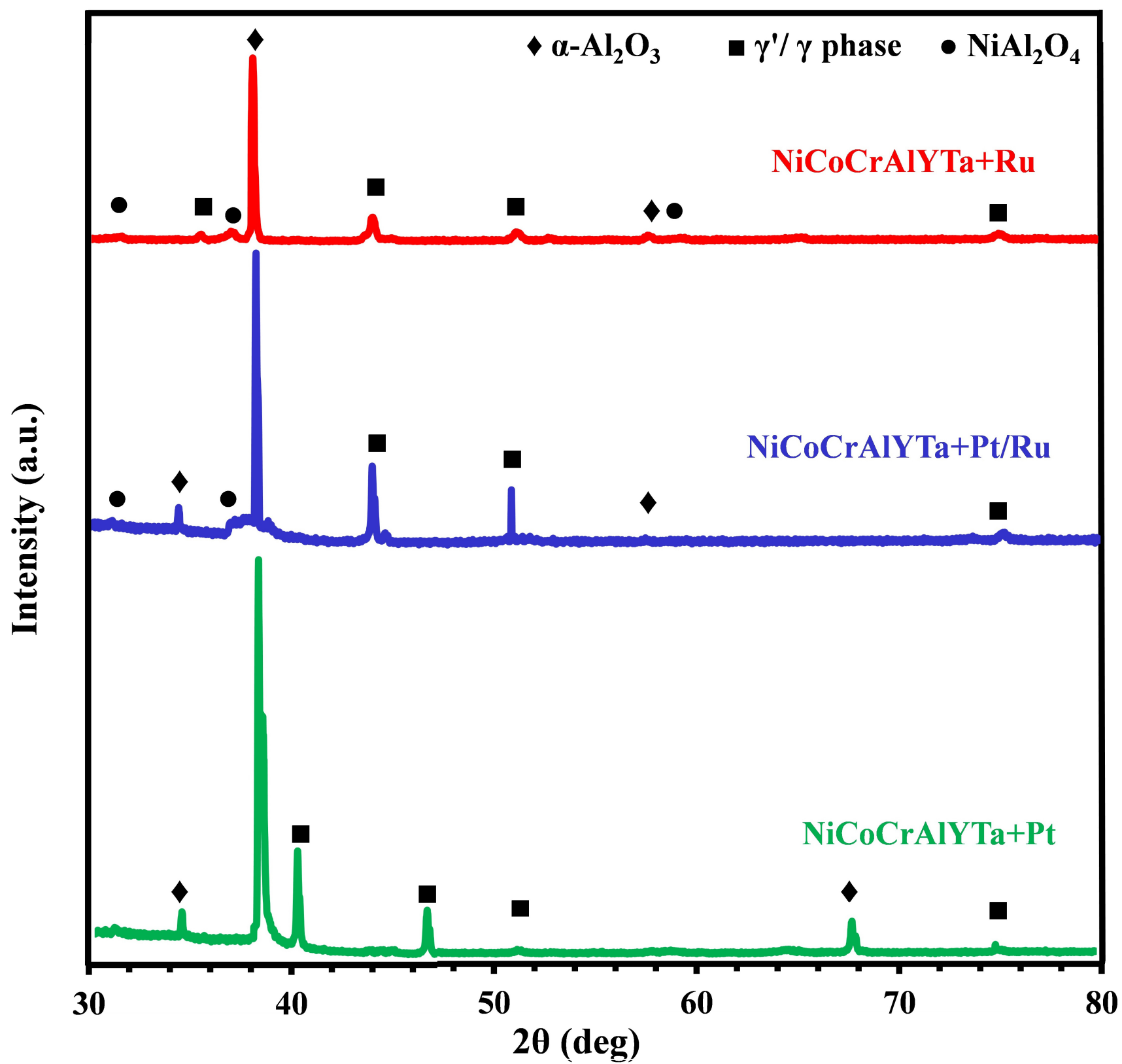

Fig. 5. XRD patterns of the NiCoCrAlYTa+Pt, NiCoCrAlYTa+Ru, and NiCoCrAlYTa+Pt/Ru samples.

## 4. Conclusions

This study demonstrated that the electrodeposition of Pt, Ru, and Pt-Ru onto NiCoCrAlTaY-coated CMSX-4 superalloy, followed by vacuum annealing at 1080 °C for 6 h, significantly improves oxidation resistance at 1100 °C. Among the three coatings, Pt provided the highest oxidation protection, followed by the Pt-Ru alloy, with Ru offering the least protection. Despite Ru's lower performance in terms of oxidation resistance, its cost-effectiveness and high-temperature properties offer potential for replacing or supplementing

Pt in cost-sensitive applications. These findings suggest that the Pt-Ru alloy could serve as a viable compromise between performance and cost, providing an intermediate solution for advanced high-temperature systems. Future work may focus on optimizing the alloy composition to further balance these factors for industrial applications.